\documentclass[preprint,12pt,number,sort&compress]{elsarticle}

\usepackage[T1]{fontenc}
\usepackage[utf8]{inputenc}
\usepackage{amsmath,amssymb}
\usepackage{graphicx}
\usepackage{booktabs}
\usepackage{xcolor}
\usepackage{setspace}

\journal{Applied Thermal Engineering}

\graphicspath{{./}}

\begin{document}

\onehalfspacing

\begin{frontmatter}

\title{Comparative assessment of film cooling strategies for thermal protection of a kerosene-fueled oblique detonation combustor}

\author[a]{Jianghong Li}
\author[a,b]{Songbai Yao\corref{cor1}}
\ead{yaosongbai@nimte.ac.cn}
\author[a,b]{Wenwu Zhang}

\affiliation[a]{organization={Zhejiang Key Laboratory of Laser Extreme Manufacturing for Difficult-to-Machine Materials, Ningbo Institute of Materials Technology and Engineering, Chinese Academy of Sciences},
            city={Ningbo},
            postcode={315201},
            country={China}}

\affiliation[b]{organization={University of Chinese Academy of Sciences},
            city={Beijing},
            postcode={100049},
            country={China}}

\cortext[cor1]{Corresponding author.}

\begin{abstract}

Thermal protection remains a major challenge for oblique detonation engines (ODEs), whose combustors combine intense heat release with a compact geometry anchored on oblique detonation waves (ODWs). This study numerically investigates the effects of coolant type and mass flow rate on the flow structure, thermal environment, and propulsion performance of a film-cooled kerosene--air ODE combustor at Mach 10 and an altitude of 15 km. Three active cooling strategies are compared under identical geometry and injection conditions: air film cooling, gaseous-kerosene film cooling, and liquid-kerosene mist cooling. Stable, wall-attached oblique-detonation propagation and the canonical wave structure are maintained over the investigated range, while the wave--coolant interaction depends systematically on the coolant medium. Air injection causes greater disturbance to the initiation region and strengthens downstream wave interactions, resulting in the greatest propulsion penalty. The cooling strategies substantially reduce the near-wall thermal load, but produce distinct wall-temperature distributions. Gaseous-kerosene cooling generates a spatially periodic response associated with the discrete injection slots/ports, whereas liquid-kerosene mist cooling produces a smoother and lower near-wall fluid temperature through enhanced two-phase mixing and phase-change heat absorption. Liquid-kerosene mist cooling provides the most favorable balance between thermal protection and propulsion performance at low to moderate injection levels. With further increases in injection level, the thermal benefit approaches saturation while the propulsion penalty continues to increase: the mean near-wall fluid temperature reduction remains approximately 87\% across both the inter-slot and cooling-slot wall segments, whereas the total specific impulse is reduced by about 3\%--5\%. These results quantify the thermal--propulsive trade-off associated with different film-cooling media and demonstrate the potential of fuel-based cooling for thermal management in hypersonic ODE combustors.

\end{abstract}

\begin{keyword}
Oblique detonation engine \sep Thermal management \sep Kerosene fuel \sep Film cooling \sep Mist cooling
\end{keyword}

\end{frontmatter}

\section{Introduction}

The advantage of the oblique detonation engine (ODE) lies in its use of the oblique detonation wave (ODW) to achieve rapid and efficient combustion at extremely high Mach numbers, which is expected to overcome the performance bottlenecks of scramjet engines and enable shorter, lighter engine structures, demonstrating significant potential in the field of hypersonic propulsion \cite{RosatoPNAS2021,VerreaultProcCombustInst2013,ViguierProcCombustInst1996}.

For example, Zhang et al. \cite{ZhangCombustFlame2022} conducted a large-scale hydrogen-fueled ODE model test in a hypersonic wind tunnel at an inflow Mach number of 6.6, and reported two stable operating modes in the combustor. In a subsequent study \cite{HanCombustFlame2023}, they performed tests on a kerosene-fueled ODE using forced initiation. Tian et al. \cite{TianJFM2024} numerically investigated the propagation instability of the ODW in pre-evaporated liquid fuel sprays and analyzed the effect of droplet diameter on its initial structure and stability modes. Sun et al. \cite{SunAIAAJ2023} studied the structural evolution of the ODW under unstable inflow conditions and proposed a control technique to enhance its stability. Ren et al. \cite{RenJFM2018} numerically studied the initiation and stabilization of the ODW with kerosene--air mixtures, finding that an increase in droplet mass flow rate changed the transition from smooth to abrupt and extended the ignition length. Guo et al. \cite{GuoCombustFlame2022} employed an Eulerian--Lagrangian framework to show that, in hydrogen--oxygen mixtures containing fine water mist, the droplet mass flux can distort the leading edge of the ODW. Wang et al. \cite{WangCombustFlame2025} reported that fuel vapor enrichment or the use of on-wedge strips could effectively promote ODW formation. Ling et al. \cite{LingAST2024} revealed that the inlet exit Mach number and combustor wedge angle showed inverted U-shaped effects on specific impulse. In a high-fidelity simulation study, Abisleiman et al. \cite{AbisleimanCombustFlame2025} elucidated the structure of three-dimensional conical ODWs. Yao et al. \cite{YaoAST2022} showed that expansion waves could re-stabilize an ODW by eliminating thermal choking and shift the initiation position downstream with decreasing wedge length.

It can be seen that current ODE research has advanced in areas such as initiation and steady-state mechanisms; however, the extreme conditions of detonation combustion make thermal protection a critical engineering challenge that must be addressed \cite{RoyPECS2004}. For example, previous studies have explored the feasibility of regenerative cooling \cite{KawalecJPP2023,JorgensenAIAA2022}, transpiration cooling \cite{LuATE2025}, and film cooling \cite{YuIJHE2024,ZhouEnergy2025} for rotating detonation engines (RDEs). In an RDE, the detonation wave propagates continuously in the circumferential direction, whereas an ODE relies on a spatially anchored oblique detonation wave initiated and stabilized by compression over a wedge. Consequently, the wedge surface in an ODE serves not only as a thermally loaded wall but also as an aerodynamic element directly involved in ODW initiation and stabilization. Coolant injection may therefore modify the oblique-shock structure, induction region, wave onset location, and downstream wave interactions. However, although the previous studies \cite{GotoJPP2022a,TianAeScTe2022a,LiAeScTe2023a,LiuApThEn2025a,RamanagarsridharaApECS2024a,YuIJHE2024,ZhouEnergy2025} demonstrate the feasibility of film cooling in rotating detonation systems, the cooling requirements and wave--coolant interaction mechanisms differ fundamentally from those of an ODE combustor. In addition, ODEs generally operate under hypersonic inflow conditions, which impose more stringent requirements on coolant selection and cooling-system integration. The severe geometric constraints associated with hypersonic ODE combustors further motivate cooling strategies with minimal intrusion into the core flow. These ODE-specific characteristics distinguish the present problem from previous film-cooling studies of RDEs, which have primarily focused on cooling configurations applicable to rotating combustor geometries. Accordingly, the present study considers two aspects that are particularly relevant to ODEs. First, the cooling system is applied directly to the wedge surface responsible for anchoring the ODW, and the effects of coolant injection on the detonation structure, near-wall thermal environment, and propulsion performance are evaluated in a coupled framework. Second, rather than considering air cooling alone, three coolant media, namely air, gaseous kerosene, and liquid-kerosene mist, are compared under the same ODE operating conditions. This framework enables a consistent assessment of air-based film cooling, gaseous-fuel film cooling, and two-phase evaporative cooling for thermal management in hypersonic ODEs.

\section{Numerical methodology and physical modeling}

\subsection{Physical and boundary conditions}

\begin{figure}[htbp]
\centering
\includegraphics[width=0.82\linewidth]{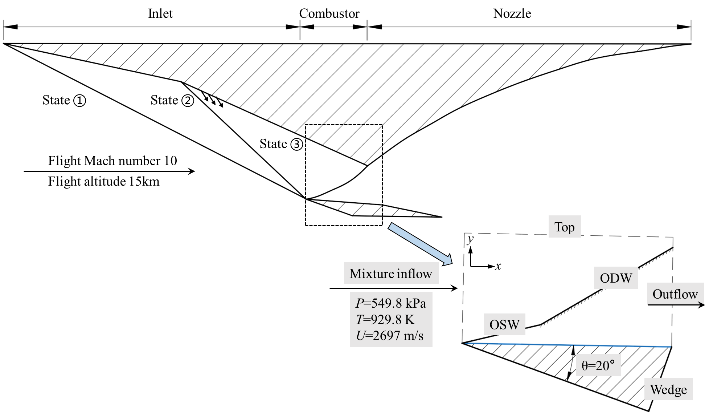}
\caption{Schematic of the oblique detonation engine configuration.}
\label{fig:config}
\end{figure}

The configuration of the ODE combustor considered in this study is shown in Fig.~\ref{fig:config}. The combustor has a height of 50 mm and a length of 40 mm, and the wedge angle is $\theta=20^\circ$ to initiate the oblique detonation. A similar configuration was previously used in Ref. \cite{TengJFM2021} to model an ODE fueled by hydrogen and air. Here, we extend this configuration to a kerosene-fueled ODE application. A two-dimensional computational domain is constructed for the region indicated by the dashed box in Fig.~\ref{fig:config}, with the wedge leading edge taken as the coordinate origin. The main flow inlet is located at the left boundary and corresponds to a flight altitude of $H_{0}=15$\,km and a flight Mach number of $M_{0}=10$, with inlet conditions $p$ = 549.8 kPa, $T$ = 929.8 K, and $U$ = 2697 m/s. The right boundary is specified as the outlet. 

The wedge surface of the ODE combustor is selected as the primary region of interest for thermal-protection analysis in the present study. The nominal cooling-hole diameter is $d=0.2$\,mm. In the two-dimensional formulation, the cooling arrangement is represented by 20 streamwise slot openings distributed along the wedge surface. Each slot has a streamwise width of approximately 0.2 mm and an approximately 2 mm streamwise pitch, and extends through the numerical span. The coolant is injected normal to the local wedge surface.
The two-dimensional model is adopted as an idealized framework to systematically isolate the effects of coolant type and phase state under identical geometric and injection conditions. The same cooling configuration is applied to air, gaseous kerosene, and liquid-kerosene mist, thereby eliminating additional geometric variations and enabling their respective effects on the ODW structure, near-wall thermal environment, and propulsion performance to be compared on a consistent basis. This idealized configuration does not resolve the lateral spreading of individual jets, spanwise jet interactions, or three-dimensional vortical structures associated with discrete circular-hole injection. Therefore, the quantitative cooling performance and relative trends reported herein should be interpreted within the present two-dimensional slot configuration, and their applicability to practical three-dimensional discrete-hole arrangements requires further investigation.

The operating condition parameters used in this study are adopted from Ref. \cite{TengJFM2021}. Air enters the inlet and is compressed through the inlet/isolator system before reaching the combustor. The fuel is kerosene, represented by the surrogate formula $\mathrm{C_{10}H_{20}}$. For all cases, the incoming stream is prescribed as a spatially uniform, stoichiometric kerosene--air mixture with an inlet equivalence ratio of $\phi=1$. The corresponding inlet mass fractions are $Y_{\mathrm{C_{10}H_{20}}}=0.0636$, $Y_{\mathrm{O_{2}}}=0.2182$, and $Y_{\mathrm{N_{2}}}=0.7182$. Under the prescribed inlet conditions, the resulting main-stream mass flow rate is approximately $\dot{m}_{\mathrm{main}}=291.4$ g/s. After being compressed by two shocks, the mixture enters the combustor at the location indicated in Fig.~\ref{fig:config}.

\begin{table}[htbp]
\centering
\caption{Summary of the ODE combustor configuration.}
\label{tab:config}
\begin{tabular}{lll}
\toprule
Parameter & Nomenclature & Values \\
\midrule
Length & $L$ [mm] & 40 \\
Height & $H$ [mm] & 50 \\
Wedge angle & $\theta$ [$^\circ$] & 20 \\
Flight Mach number & $M_{0}$ & 10 \\
Flight altitude & $H_{0}$ [km] & 15 \\
Pressure & $p$ [kPa] & 549.8 \\
Temperature & $T$ [K] & 929.8 \\
Velocity & $U$ [m/s] & 2697 \\
\bottomrule
\end{tabular}
\end{table}

Since this study focuses on film cooling, the kerosene--air mixture at the inlet is assumed to be fully gaseous. For the coolant, kerosene is considered in both liquid and gaseous states. Table~\ref{tab:config} provides a full summary of the ODE combustor geometry and inlet conditions.

Three film-cooling strategies are considered: air film cooling, gaseous-kerosene film cooling, and liquid-kerosene mist cooling. For air film cooling, the coolant is assumed to be supplied from the same inlet system and internal air passages, and its inlet temperature is therefore set equal to that of the main flow, $T_{c,\mathrm{in}}=929.8\ \mathrm{K}$. For gaseous-kerosene film cooling, the coolant is treated as fully vaporized kerosene, with an inlet temperature of approximately $T_{c,\mathrm{in}}=700\ \mathrm{K}$. In contrast, mist cooling involves the direct injection of liquid-kerosene droplets through the cooling openings at $T_{c,\mathrm{in}}=300\ \mathrm{K}$. The injection intensity is characterized directly by the prescribed coolant mass flow rate, $\dot{m}_{c}$. Four coolant mass flow rates, 0.001, 0.003, 0.009, and 0.015 kg/s, are considered for each cooling medium. At each injection level, the same coolant mass flow rate is applied to the air, gaseous-kerosene, and liquid-kerosene cases, while the main-stream inflow condition remains unchanged.

For the air- and gaseous-kerosene cooling cases, the total coolant mass flow rate is prescribed over the cooling\_inlet boundary using the flowRateInletVelocity condition. The coolant velocity is therefore determined according to the prescribed mass flow rate, local density, and inlet geometry rather than specified independently. A zero-gradient pressure condition is applied, while the coolant temperature is fixed at 929.8 K for air and 700 K for gaseous kerosene. The resulting injection velocity is normal to the local wedge surface.

For the mist-cooling cases, liquid kerosene droplets are introduced through a Lagrangian patchInjection model. The droplets are monodisperse, with an initial diameter of 5\,$\mu$m and an initial temperature of 300 K. Their mass flow rate and initial velocity are prescribed through the injection model, with the initial droplet velocity selected based on the gaseous-kerosene injection velocity at the corresponding coolant mass-flow condition. At the corresponding Eulerian gas-phase boundary, velocity is treated as noSlip, while pressure and temperature employ zero-gradient conditions.

\begin{table}[htbp]
\centering
\caption{Summary of the operating conditions.}
\label{tab:conditions}
\begin{tabular}{llll}
\toprule
Case & Coolant & $T_{c,\mathrm{in}}$ [K] & $\dot{m}_{c}$ [kg/s] \\
\midrule
A0 & / & / & / \\
A1 & Air & 929.8 & 0.001 \\
A2 & Air & 929.8 & 0.003 \\
A3 & Air & 929.8 & 0.009 \\
A4 & Air & 929.8 & 0.015 \\
G1 & Kerosene (g) & 700 & 0.001 \\
G2 & Kerosene (g) & 700 & 0.003 \\
G3 & Kerosene (g) & 700 & 0.009 \\
G4 & Kerosene (g) & 700 & 0.015 \\
L1 & Kerosene (l) & 300 & 0.001 \\
L2 & Kerosene (l) & 300 & 0.003 \\
L3 & Kerosene (l) & 300 & 0.009 \\
L4 & Kerosene (l) & 300 & 0.015 \\
\bottomrule
\end{tabular}
\end{table}

Table~\ref{tab:conditions} lists the operating conditions investigated. Using the uncooled case A0 as the baseline, three cooling strategies are considered under the same main-flow condition: air film cooling (Cases A1--A4, $T_{c,\mathrm{in}}$ = 929.8 K), gaseous-kerosene film cooling (Cases G1--G4, $T_{c,\mathrm{in}}$ = 700 K), and liquid-kerosene mist cooling (Cases L1--L4, $T_{c,\mathrm{in}}$ = 300 K). For all cases, the main inflow is maintained at the same prescribed inlet pressure, temperature, velocity, and stoichiometric kerosene--air composition. The coolant is introduced as an additional stream through the cooling openings; consequently, Cases G and L involve additional kerosene supplied locally to the near-wall region, while the stoichiometric core inflow remains unchanged.

For the continuous air and gaseous-kerosene injections, the inlet-referenced blowing ratio $M_{c}$ and momentum-flux ratio $J$ are additionally evaluated as

\begin{equation}
M_{c}=\frac{(\rho_{c}U_{c,n})_{A}}{\rho_{m}U_{m}}
\label{eq:blowing}
\end{equation}

\begin{equation}
J=\frac{(\rho_{c}U_{c,n}^{2})_{A}}{\rho_{m}U_{m}^{2}}
\label{eq:momentum}
\end{equation}
where $U_{c,n}$ is the coolant velocity component normal to the cooling surface, and $(\cdot)_{A}$ denotes area averaging over the cooling-inlet boundary. Here, $\rho_{m}$ and $U_{m}$ denote the mainstream density and velocity at the main-flow inlet, which is used as the reference state. The corresponding values are listed in Table~\ref{tab:blowing}. The reported $M_{c}$ and $J$ values are time-averaged over $t=0.2$--$0.3$\,ms. The parameters of $M_{c}$ and $J$ are not assigned to the mist-cooling cases because the liquid phase is represented by dispersed Lagrangian droplets rather than a continuous liquid jet.

\begin{table}[htbp]
\centering
\caption{Inlet-referenced blowing and momentum-flux ratios for the cooling cases.}
\label{tab:blowing}
\begin{tabular}{lll}
\toprule
Case & $M_{c}$ & $J$ \\
\midrule
A1 & 0.0429 & $2.32\times10^{-4}$ \\
A2 & 0.1287 & $2.09\times10^{-3}$ \\
A3 & 0.3861 & $1.57\times10^{-2}$ \\
A4 & 0.6435 & $4.07\times10^{-2}$ \\
G1 & 0.0429 & $3.62\times10^{-5}$ \\
G2 & 0.1287 & $3.06\times10^{-4}$ \\
G3 & 0.3861 & $2.58\times10^{-3}$ \\
G4 & 0.6435 & $7.14\times10^{-3}$ \\
L1--L4 & - & - \\
\bottomrule
\end{tabular}
\end{table}

\subsection{Numerical methods}

The simulations are conducted using a modified solver based on OpenFOAM. The solver has been used and validated in our previous studies for RDE simulations \cite{LiActaAstronautica2023,LiPhysFluids2024,LiIJHE2025}. For the two-phase detonation, this study employs a point-source approximation within an Eulerian--Lagrangian framework. The gas phase is solved using the compressible Navier--Stokes (N-S) equations with source terms coupled to the droplets, while the liquid phase is tracked in a Lagrangian framework to describe droplet motion and evaporation. The evaporation of the Lagrangian droplets is modeled by assuming ideal-gas behavior for the surface vapor and vapor--liquid equilibrium at the droplet surface. When the saturation pressure at the droplet surface approaches the local pressure, boiling is treated by an energy-limited formulation with the droplet temperature capped at the local boiling temperature. The evaporation rate is computed via a convective-diffusive formulation. Two-way coupling between droplets and the gas phase is implemented through source terms for mass, momentum, energy, and species. \begin{equation}
\dot{m}=\pi d_{p}\,\mathrm{Sh}\,D_{v}\rho_{v}\ln\left(1+B_{M}\right)
\label{eq:evaporation}
\end{equation}
where $\mathrm{Sh}=2+0.6\mathrm{Re}^{1/2}\mathrm{Sc}^{1/3}$ is the Sherwood number given by the Ranz-Marshall correlation \cite{RanzChemEngProg1952b,RanzChemEngProg1952a}, $D_{v}$ is the vapor diffusivity, $\rho_{v}$ is the vapor density at the droplet surface evaluated from the ideal gas law, and $B_{M}=(X_{s}-X_{c})/(1-X_{s})$ is the molar-fraction-based Spalding mass transfer number; $X_{s}$ is the surface mole fraction from Raoult's law and $X_{c}$ is the carrier-phase mole fraction.

\begin{figure}[htbp]
\centering
\includegraphics[width=0.55\linewidth]{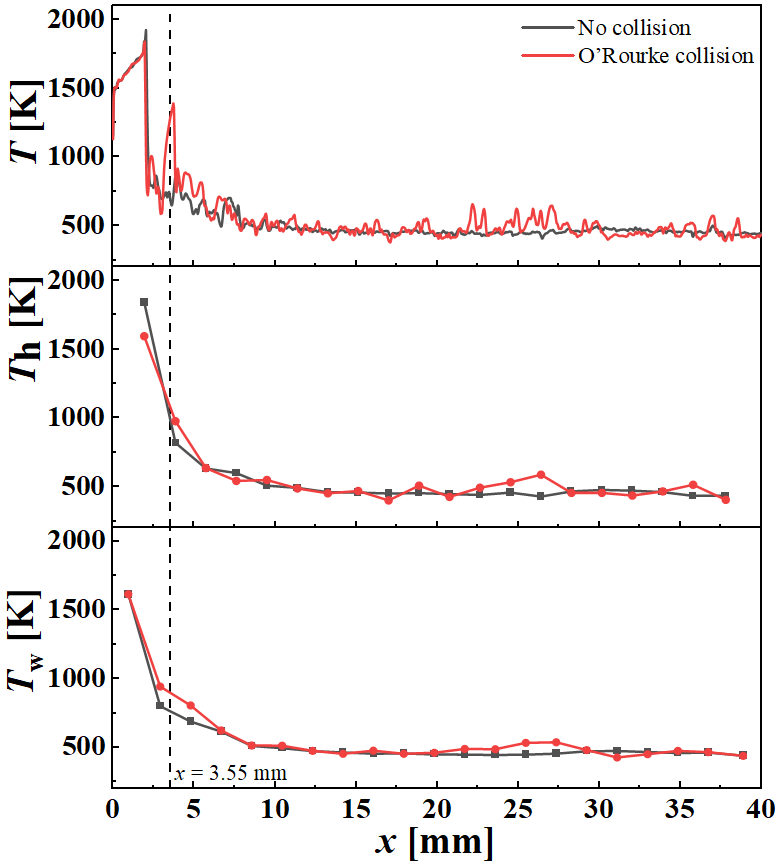}
\caption{Sensitivity of the near-wall temperature statistics to the droplet--droplet collision model.}
\label{fig:collision}
\end{figure}

According to Ref. \cite{ZhaoArXiv2020}, the secondary breakup has a negligible influence when the initial droplet diameter satisfies $d_{p}<20\,\mu$m for the Reitz KH--RT breakup model, whereas only minor differences appear when $d_{p}$ increases to about 80\,$\mu$m. In the present simulations, the injected droplets have an initial diameter of $d_{p}=5\,\mu$m. Additionally, a sensitivity calculation using the O'Rourke collision model with coalescence disabled shows that the mean $\bar{T}_{h}$ and $\bar{T}_{w}$ differ from the no-collision case by 4.31\% and 4.46\%, respectively, indicating a limited influence of droplet--droplet collisions on the overall near-wall thermal response (Fig.~\ref{fig:collision}).

For the kerosene--air combustion, a two-step mechanism proposed by Franzelli et al. \cite{FranzelliCombustFlame2010} is used:

\begin{equation}
\mathrm{C_{10}H_{20}}+10\,\mathrm{O_{2}}\Rightarrow10\,\mathrm{CO}+10\,\mathrm{H_{2}O}
\label{eq:reaction-1}
\end{equation}

\begin{equation}
\mathrm{CO}+0.5\,\mathrm{O_{2}}\Leftrightarrow\mathrm{CO_{2}}
\label{eq:reaction-2}
\end{equation}
This two-step mechanism has been extensively employed and validated in kerosene-fueled detonation simulations \cite{WangFTC2023,SalvadoriAIAAJ2023,ZhangJPP2025}.

Spatial discretization is performed using the Kurganov-Tadmor (K-T) central-upwind scheme, which is widely adopted in OpenFOAM-based density-based solvers (e.g., rhoCentralFoam and its derivatives) for robust shock capturing in high-speed reactive flows. Time integration is handled with a second-order implicit backward scheme, with an adaptive time step to ensure the maximum Courant--Friedrichs--Lewy (CFL) number remains below 0.1. The dynamic viscosity $\mu$ is calculated using Sutherland's law, i.e., $\mu=\frac{A_{s}\sqrt{T}}{1+T_{s}/T}$, where $A_{s}$ is set to $1.672\times10^{-6}\ \mathrm{kg/(m{\cdot}s{\cdot}\sqrt{K})}$, and $T_{s}=170.672$ K. Under the assumption that the Lewis number is unity, the species mass diffusivity, $D_{m}$, is computed from the thermal diffusivity as $D_{m}=k/(\rho C_{p})$, where the thermal conductivity $k$ is determined using the Eucken approximation, and $C_{p}$ is the heat capacity.

Following Ref. \cite{FigueiraDaSilvaCombustFlame2000}, a no-slip, adiabatic wall boundary condition is implemented for the wedge. The turbulence is treated using an implicit large eddy simulation (iLES) approach adopted by Desai et al. \cite{DesaiCombustFlame2023}. In accordance with this approach, the near-wall mesh is refined. Statistics show that approximately 90\% of sampling points satisfy $y^{+}<5$, indicating that most data lie within the viscous sublayer and reliably represent the wall-adjacent thermal environment. The minimum near-wall cell size is approximately 5\,$\mu$m, the normal growth ratio is limited to 1.15, and the maximum cell size in the domain is constrained to 40\,$\mu$m.

\subsection{Model validations}
\begin{figure}[!htbp]
\centering
\includegraphics[width=0.65\linewidth]{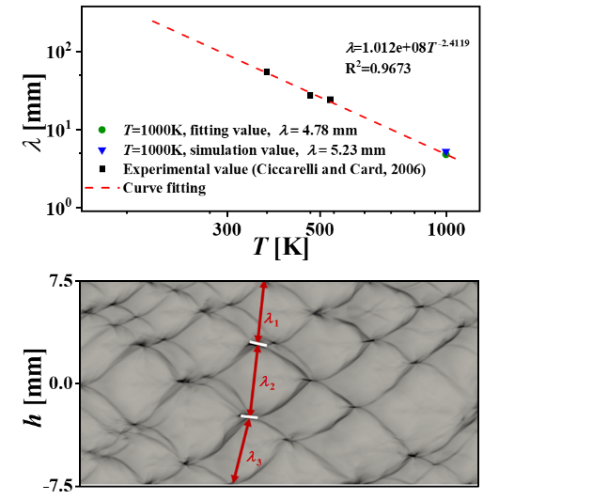}
\caption{Comparison of the predicted detonation cell size with the JP-10 experimental correlation of Ciccarelli and Card \cite{CiccarelliAIAAJ2006}.}
\label{fig:cell-size}
\end{figure}

The reduced combustion mechanism is validated first, followed by the numerical scheme, the film-cooling treatment, and a mesh-sensitivity assessment. Following Refs. \cite{LingAST2024,LiuFuel2021}, its detonation-scale behavior is assessed using the present solver by comparing the predicted detonation cell size with the experimental data of Ciccarelli and Card \cite{CiccarelliAIAAJ2006}, as shown in Fig.~\ref{fig:cell-size}. The experimental correlation is based on JP-10, rather than the kerosene surrogate ($\mathrm{C_{10}H_{20}}$) used in the present simulations. Therefore, the comparison is intended only as a detonation-scale assessment.

\begin{figure}[!htbp]
\centering
\includegraphics[width=0.85\linewidth]{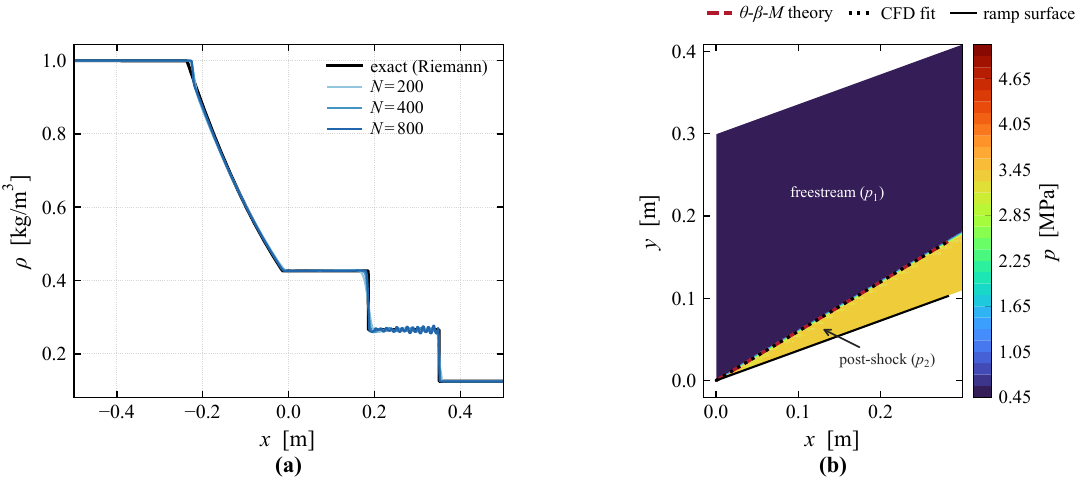}
\caption{Verification of the density-based central-upwind scheme for non-reacting shocks.}
\label{fig:scheme_verification}
\end{figure}

The density-based central-upwind scheme used throughout this study is verified against two benchmark problems with exact solutions, as shown in Fig.~\ref{fig:scheme_verification}. For the Sod shock tube at $t=6.3213\times10^{-4}$~s, the relative $L_1$ density error decreases monotonically from 0.3152\% to 0.2221\% and 0.1904\% on meshes with $N=200$, 400, and 800 cells, respectively. The corresponding pressure errors decrease from 0.2869\% to 0.2346\%, while the computed shock-front position remains within 0.09\% of the exact solution at all resolutions. 

The scheme is further verified for a non-reacting $20^\circ$ compression ramp using the same static pressure and temperature as the combustor inflow ($T=929.8$~K and $p=5.498\times10^{5}$~Pa) at an inflow Mach number of $M=4.5$. The numerically obtained shock angle agrees with the exact $\theta$--$\beta$--$M$ relation to within 0.24\% on all three meshes containing 5,000, 20,000, and 80,000 cells. The corresponding post-shock pressure-ratio error decreases from 0.50\% to 0.015\% from the coarsest to the finest mesh, while the post-shock temperature-ratio error decreases from 1.97\% to 1.72\% and 0.37\%. These results demonstrate the accuracy of the computed shock location and post-shock state under the conditions relevant to the present ODW simulations.

In addition to the baseline compressible-flow validation, the numerical treatment of film cooling is further assessed against the supersonic air-film-cooling experiments of Juhany et al. \cite{JuhanyJTHT1994}. The experiments were conducted at a freestream Mach number of $M_{\infty}=2.44$ with a coolant-slot height of $s=1.5$\,mm and a freestream total temperature of approximately 295 K. Three cooled-air injection conditions from the experiments were considered, corresponding to $M_{i}$ = 1.2, 1.5, and 1.8. The corresponding coolant-to-freestream total-temperature ratios $T_{ti}/T_{t\infty}$ were 0.80, 0.76, and 0.76; the velocity ratios $U_{i}/U_{\infty}$ were 0.57, 0.66, and 0.73; and the mass-flux ratios $\lambda$ were 0.40, 0.59, and 0.74, respectively.

The mainstream and coolant conditions in the numerical model are prescribed according to these experimental conditions, with the coolant static pressure matched to that of the freestream. The experimentally reported turbulent boundary-layer thickness of 3.4 mm at the beginning of the injection slot is also reproduced at the numerical inlet. Film-cooling effectiveness was evaluated following the definition adopted by Juhany et al. \cite{JuhanyJTHT1994}.

\begin{figure}[!htbp]
\centering
\includegraphics[width=0.55\linewidth]{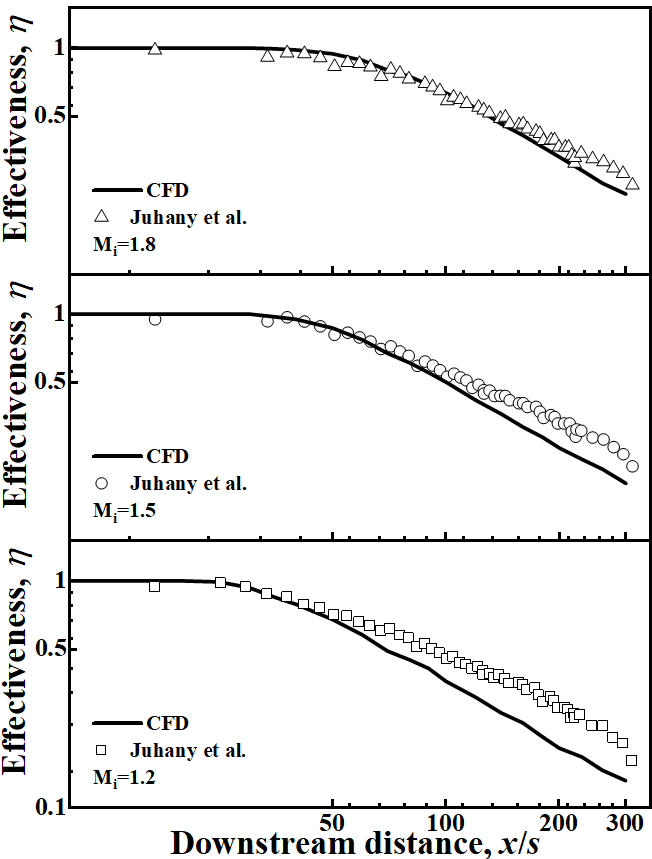}
\caption{Comparison of the predicted film-cooling effectiveness with the experimental data of Juhany et al. \cite{JuhanyJTHT1994} for cooled-air injection at $M_{i}$ = 1.2, 1.5, and 1.8.}
\label{fig:fcool}
\end{figure}

Fig.~\ref{fig:fcool} compares the predicted film-cooling effectiveness with the experimental measurements for the three injectant Mach numbers. The calculations reproduce the overall streamwise decay of film-cooling effectiveness and its variation with injectant Mach number. The mean absolute errors in $\eta$ are 0.087, 0.057, and 0.035 for $M_{i}$ = 1.2, 1.5, and 1.8, respectively. This comparison provides additional validation of the numerical treatment of supersonic film cooling and the associated near-wall thermal behavior.

\begin{figure}[htbp]
\centering
\includegraphics[width=0.55\linewidth]{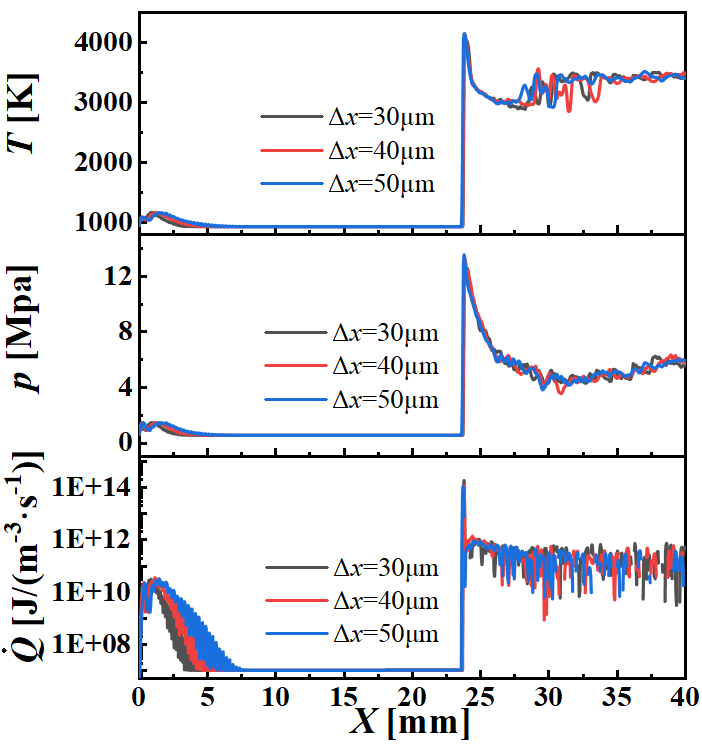}
\caption{Distributions of temperature, pressure, and heat-release rate along the line passing through initiation and triple point.}
\label{fig:mesh}
\end{figure}

Meanwhile, a mesh sensitivity analysis is conducted to assess grid dependence using three different mesh sizes of 30\,$\mu$m, 40\,$\mu$m, and 50\,$\mu$m. Fig.~\ref{fig:mesh} shows the distributions of temperature, pressure, and heat release rate along a line traversing both the initiation points and the triple point for the three mesh resolutions. The results indicate that all three meshes effectively capture the key structural features of the oblique detonation flow, including the triple-point location, wave-front propagation, and reflected-wave system.

To further quantify the mesh sensitivity, the peak pressure, peak temperature, and ODW onset location are extracted for the three meshes. For mesh sizes of 30, 40, and 50\,$\mu$m, the peak pressures are 12.567, 13.350, and 13.560 MPa, respectively, while the corresponding peak temperatures are 4035.0, 4142.5, and 4142.5 K. The predicted ODW onset locations are 23.68, 23.68, and 23.60 mm, respectively, with a maximum difference of only 0.08 mm. In particular, compared with the 30\,$\mu$m mesh, the adopted 40\,$\mu$m mesh predicts an identical ODW onset location, while the differences in peak pressure and peak temperature are 6.23\% and 2.66\%, respectively.

Considering the consistent prediction of the characteristic ODW structure and the principal near-wall thermal response, together with the balance between computational accuracy and cost, the 40\,$\mu$m mesh is selected for all subsequent simulations. It should be noted, however, that the near-wall mesh refinement required to satisfy $y^{+}<5$ yields a minimum cell size of approximately $5\,\mu$m; at this scale the grid-to-droplet size ratio approaches unity and the point-source assumption of the Lagrangian treatment is locally violated. The initial droplet diameter $d_{p}=5\,\mu$m is small compared with the characteristic mesh size $\Delta x=40\,\mu$m in the main flow region ($d_{p}/\Delta x=0.125$), and the leading-order error introduced by representing the finite-size droplet source as a point source is of order $(d_{p}/\Delta x)^{2}\approx1.6\%$. The Knudsen number evaluated at the post-shock state at the cooling station ($p_{2}=3.34$ MPa, $T_{2}=1822$ K, $U_{2}=2402$ m/s) is $\mathrm{Kn}=\ell/d_{p}\approx3.2\times10^{-3}$, with the molecular mean free path $\ell\approx16$ nm, well below the continuum limit of $0.01$. The liquid volume fraction is largest at the slot exit and decreases further downstream through mixing and evaporation, so that the spray remains in the dilute regime. Finally, the droplet relaxation time $\tau_{p}=\rho_{l}d_{p}^{2}/(18\mu_{g})=15.4\,\mu$s, with $\rho_{l}$ evaluated from the liquid-property correlation at the injection temperature and $\mu_{g}$ the gas viscosity from Sutherland's law at the post-shock temperature of the cooling station, is nearly three orders of magnitude larger than the cell-crossing time $t_{\mathrm{cross}}=\Delta x/U_{2}\approx1.7\times10^{-8}$ s, corresponding to a Stokes number $\mathrm{St}=\tau_{p}/t_{\mathrm{cross}}\approx9.3\times10^{2}$; the droplets are therefore strongly inertial and do not follow the smallest resolved scales, so that their near-wall transport is governed primarily by large-scale convection rather than by the unresolved small-scale momentum exchange in the wall-adjacent cells; this is inherent to the parcel-based treatment of micron-scale sprays in supersonic crossflow.

\section{Results and discussion}

This section first discusses the baseline flow-field and wall-thermal characteristics of the uncooled ODE combustor. The effects of the three cooling strategies are then examined in terms of detonation-wave structure and propagation, near-wall thermal response and cooling performance, and finally propulsion performance.

\subsection{Baseline flow field and wall thermal load characteristics}

\begin{figure}[htbp]
\centering
\includegraphics[width=0.88\linewidth]{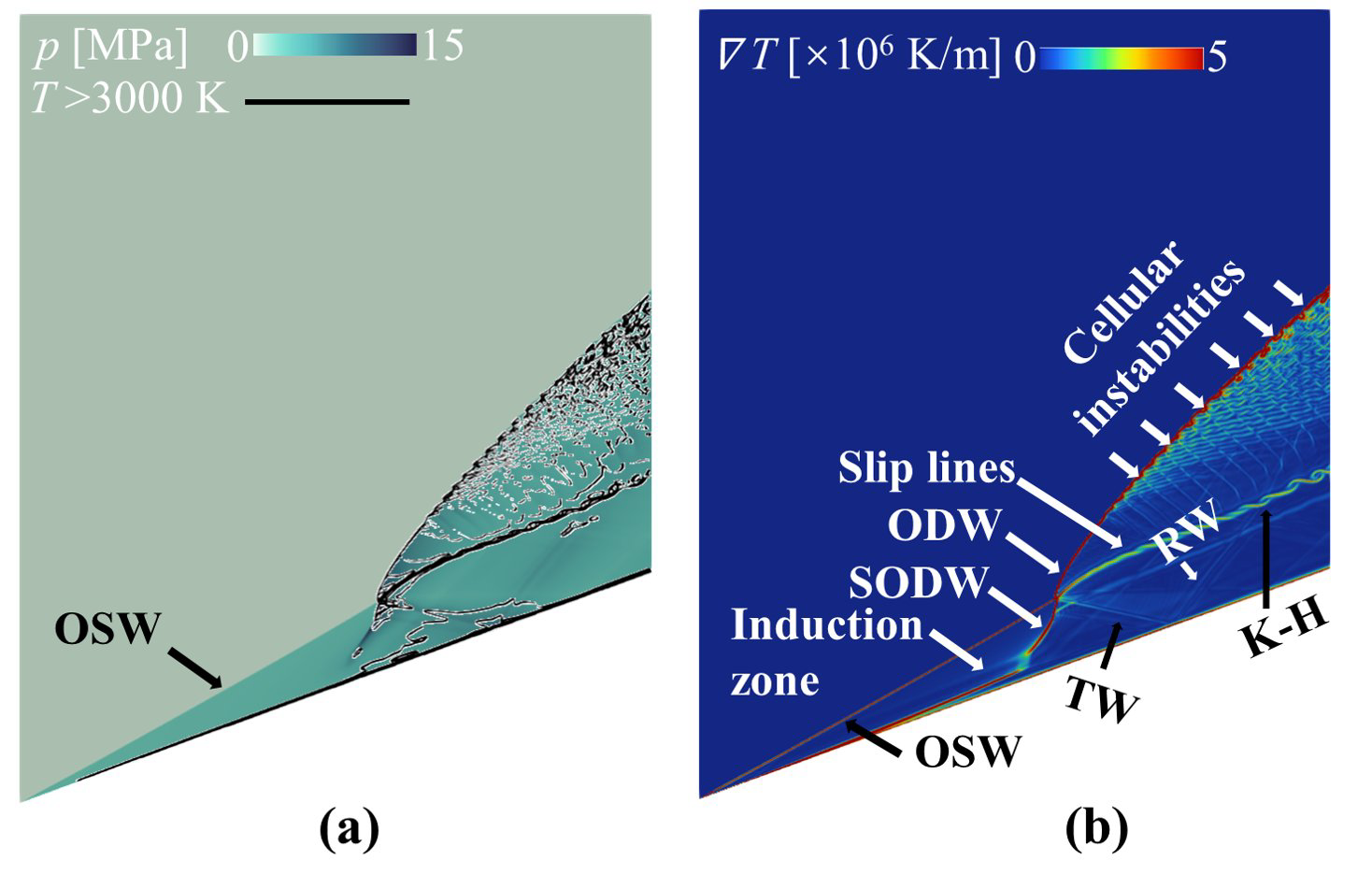}
\caption{Pressure (a) and temperature gradient magnitude (b) fields of the baseline case (Case A0).}
\label{fig:baseline}
\end{figure}

Fig.~\ref{fig:baseline} presents the baseline flow field in the ODE combustor without cooling. The pressure field and temperature gradient contours reveal a canonical ODW system developing along the inclined wall. Owing to the strong compression induced by the wedge geometry, a non-reacting oblique shock wave forms near the inlet. Downstream of this shock, a near-wall induction region is established, where the mixture undergoes continued compression and heating and exothermic reactions are first triggered; the near-wall temperature can exceed 3000 K. As the heat release progressively couples with the oblique shock, a shock-induced secondary oblique detonation wave develops and, after further compression, transitions into a stabilized oblique detonation wave attached to the wall.

At sufficient resolution, the ODW front evolves from an initially smooth segment into a cellular pattern downstream; this region is labeled as cellular instabilities in Fig.~\ref{fig:baseline}, consistent with prior observations of the instability of the ODW front \cite{PapalexandrisCombustFlame2000}. The post-wave slip line also becomes unstable and develops characteristic Kelvin-Helmholtz (K-H) vortical structures farther downstream. Transverse waves and their wall-induced reflections can be identified in the post-detonation region, but these small-scale features do not alter the macroscopic ODW topology.

Fig.~\ref{fig:baseline}(a) overlays the pressure field with the isotherm ($T\geq 3000$ K) contour. The detonation wave remains stabilized in a wall-attached manner with an essentially fixed anchoring location, while high-temperature, high-pressure products sweep along the wall. The wall is therefore directly exposed to hot products over a long streamwise distance. The combined effects of shock compression, rapid chemical heat release, and high-speed near-wall transport produce a severe thermal environment, underscoring the need for effective wall thermal protection strategies under such conditions.

\subsection{Effects of different cooling strategies on the flow-field structure}

\begin{figure}[htbp]
\centering
\includegraphics[width=0.62\linewidth]{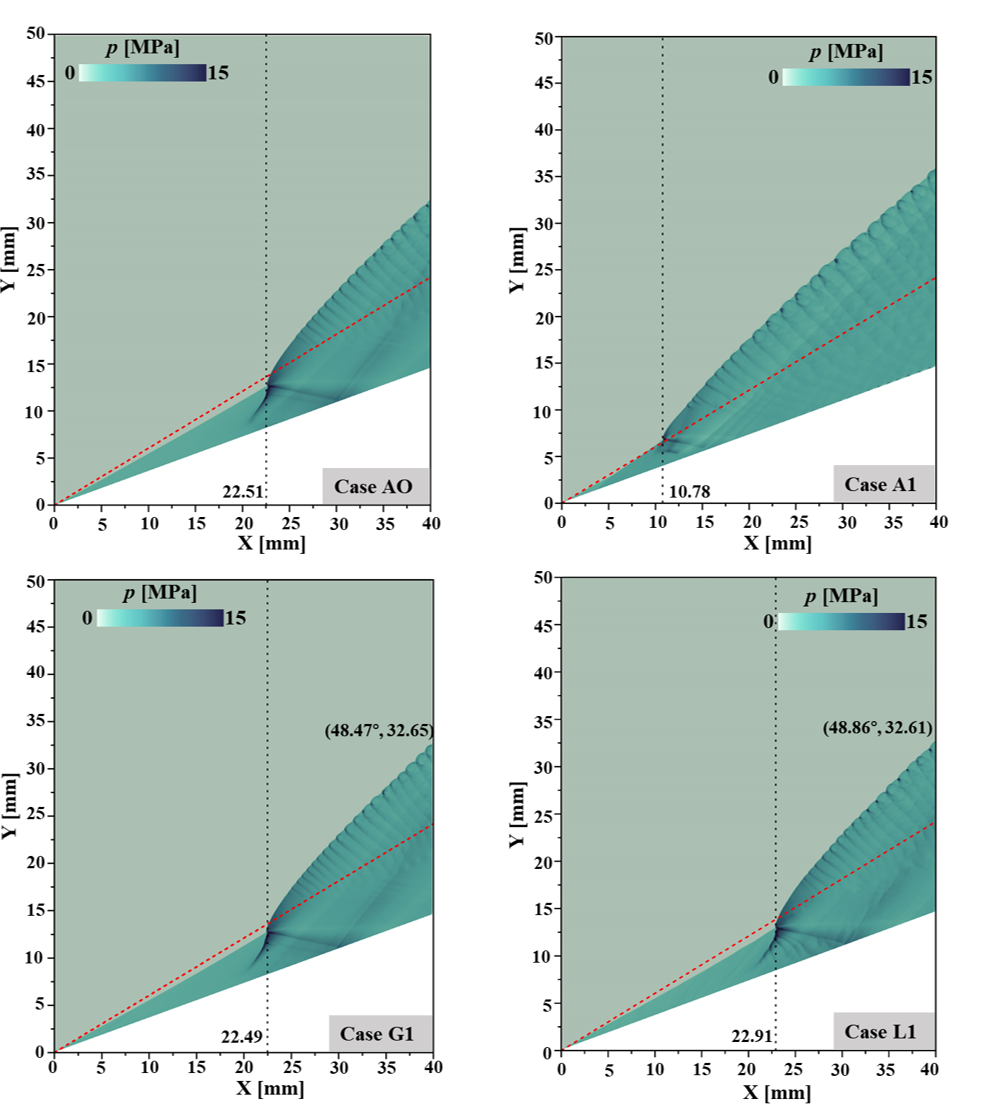}
\caption{Comparison of pressure fields and wave structures for the uncooled baseline (Case A0) and different cooling strategies at $\dot{m}_{c}=0.001$\,kg/s. The red dashed lines indicate the theoretical oblique-shock positions, while the black dotted lines denote the maximum streamwise extent of the wave front, $x_{\mathrm{max}}$.}
\label{fig:pressure-coolants}
\end{figure}

To characterize the initiation region of the oblique detonation, the wave onset location is quantified. A reference shock angle is first obtained from the classical $\theta$--$\beta$--$M$ relation for the prescribed inflow and wedge angle $\theta$ using $\gamma=1.4$, yielding $\beta=31.17^\circ$. In Fig.~\ref{fig:pressure-coolants}, the corresponding theoretical oblique-shock line is shown as a red dashed line and is used as a reference for the detonation wave inclination and onset region in the reacting flow. Fig.~\ref{fig:pressure-coolants} compares the wave-system structure for the uncooled baseline, Case A0, and the three cooling strategies at $\dot{m}_{c}=0.001$\,kg/s, namely Cases A1, G1, and L1. In all cases, the computed detonation wave inclination is slightly smaller than the theoretical oblique shock angle, indicating that exothermicity feeds back on the compression structure and causes deviations from the ideal non-reacting oblique shock model.

Among the cooled cases, A1 exhibits the closest inclination to the theoretical value, suggesting a compression field that is more dominated by gas-dynamic effects. To quantify the streamwise position of the wave front, the pressure gradient magnitude is defined as $G(x,y)=\left|\nabla p(x,y)\right|$. On a given cross-stream section at fixed $y$, the wave-front location is identified by the streamwise coordinate of the primary peak of $G$, namely $x_{f}(y)=\underset{x\in\Omega_{f}}{\arg\max}\,G(x,y)$, where $\Omega_{f}$ encloses the main detonation wave system to avoid misidentification of reflected shocks or secondary waves. The maximum streamwise extent of the wave front is then defined as $x_{\mathrm{max}}=\max_{y}x_{f}(y)$, and is indicated in Fig.~\ref{fig:pressure-coolants} by the black dotted line $x=x_{\mathrm{max}}$.

The extracted $x_{\mathrm{max}}$ values for all cases are summarized in Table~\ref{tab:xmax}. The results show distinct responses of the wave-front position to the three cooling strategies. Air cooling induces a pronounced upstream displacement even at low coolant injection, with $x_{\mathrm{max}}$ decreasing from 10.78 mm in Case A1 to 2.59 mm in Case A4 as the coolant injection increases. Gaseous-kerosene cooling causes substantially weaker displacement at low injection levels, with $x_{\mathrm{max}}=22.49$ and 22.32 mm for Cases G1 and G2, respectively, close to the uncooled value of 22.51 mm; however, the displacement becomes increasingly pronounced at higher injection levels, with $x_{\mathrm{max}}$ decreasing to 16.31 and 10.74 mm for Cases G3 and G4, respectively. In contrast, liquid-kerosene mist cooling maintains the wave-front position close to the uncooled baseline over the entire investigated injection range, with $x_{\mathrm{max}}$ varying only between 21.79 and 23.29 mm. Overall, the upstream displacement of the wave front is strongest for air cooling and increases continuously with coolant injection, while gaseous-kerosene cooling exhibits an intermediate response that becomes significant only at higher injection levels. Liquid-kerosene mist cooling produces the weakest perturbation, maintaining a wave-front position close to the uncooled baseline throughout the investigated range.

\begin{table}[htbp]
\centering
\caption{Maximum streamwise extent of the wave front for all cases.}
\label{tab:xmax}
\begin{tabular}{llllll}
\toprule
Case & $x_{\mathrm{max}}$ [mm] & Case & $x_{\mathrm{max}}$ [mm] & Case & $x_{\mathrm{max}}$ [mm] \\
\midrule
A0 & 22.51 & - & - & - & - \\
A1 & 10.78 & G1 & 22.49 & L1 & 22.91 \\
A2 & 8.41 & G2 & 22.32 & L2 & 22.77 \\
A3 & 3.95 & G3 & 16.31 & L3 & 23.29 \\
A4 & 2.59 & G4 & 10.74 & L4 & 21.79 \\
\bottomrule
\end{tabular}
\end{table}

In addition to the upstream displacement, Fig.~\ref{fig:pressure-coolants} shows that Case A1 exhibits a more pronounced deformation of the downstream wave structure, whereas Cases G1 and L1 retain wave patterns more similar to the uncooled baseline.

\begin{figure}[htbp]
\centering
\includegraphics[width=0.65\linewidth]{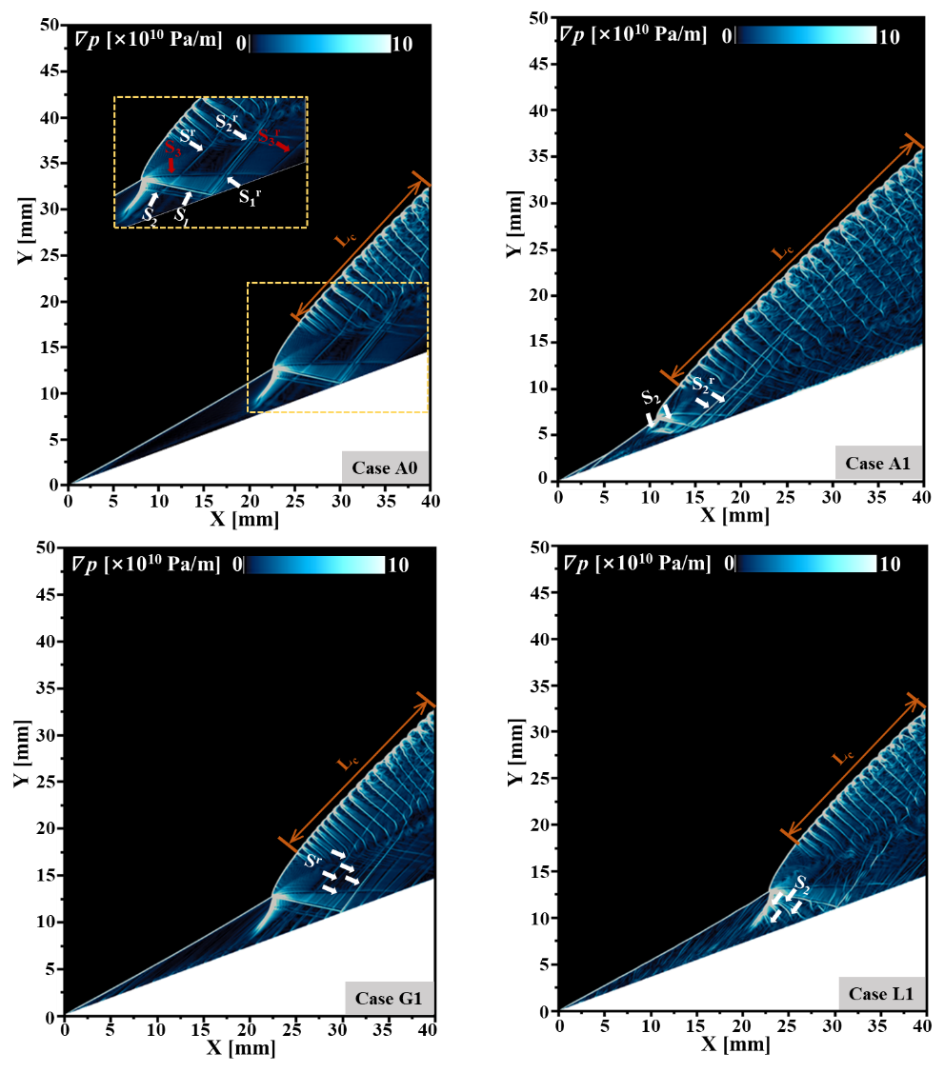}
\caption{Pressure gradient magnitudes of the cases with different coolants at $\dot{m}_{c}=0.001$\,kg/s.}
\label{fig:pgrad-coolants}
\end{figure}

Overall, coolant injection does not eliminate the ODW system, but the coolant medium can substantially modulate the wave inclination, onset location, and downstream propagation path, providing a basis for subsequent analyses of wall thermal loads and propulsion performance.

\begin{table}[htbp]
\centering
\caption{Normalized cellular extent $\Lambda_{c}$ for all cooling cases, with $L_{c,A_{0}}$ = 21.13 mm for the uncooled baseline.}
\label{tab:lambda}
\begin{tabular}{llllll}
\toprule
Case & $\Lambda_{c}$ & Case & $\Lambda_{c}$ & Case & $\Lambda_{c}$ \\
\midrule
A1 & 1.44 & G1 & 0.83 & L1 & 0.80 \\
A2 & 1.58 & G2 & 0.91 & L2 & 1.07 \\
A3 & 1.85 & G3 & 1.14 & L3 & 0.78 \\
A4 & 1.90 & G4 & 1.42 & L4 & 0.84 \\
\bottomrule
\end{tabular}
\end{table}

Using the uncooled baseline, Case A0, as an example, the zoomed view in Fig.~\ref{fig:pgrad-coolants} highlights a canonical wave system in the vicinity of the detonation. It includes an interaction shock $S_{1}$ generated by the early-stage interaction between the detonation and the incident oblique shock, together with its wall reflected counterpart $S_{1}^{r}$; a precursor shock $S_{2}$ induced by the detonation leading edge and its reflection $S_{2}^{r}$; and a strengthened shock $S_{3}$ produced as the incident oblique shock is processed by the detonation structure, along with its reflection $S_{3}^{r}$. Near the wall, an additional family of secondary shocks $S_{r}$ is also observed, triggered by localized strong compression and secondary heat release. Downstream of the detonation, the slip line separates the flow into an upper and a lower region, and cellular instabilities are observed farther downstream on the detonation front. To quantify the extent of the cellular region, the orange double-headed arrow defines $L_{c}$ as the distance measured along the main detonation front direction from the onset of cellularity to the outlet boundary. $L_{c}$ is used as a metric for comparing the cellular development among different cases.

Across the three cooled cases shown in Fig.~\ref{fig:pgrad-coolants}, the overall wave-system topology remains similar, but the wave intensity and the extent of cellular development differ markedly. For air film cooling, Case A1, the precursor shock $S_{2}$ associated with the detonation leading edge and its wall reflection $S_{2}^{r}$ are more pronounced. The onset of cellularity shifts upstream and the smooth front segment shortens, resulting in a larger $L_{c}$. For gaseous-kerosene film cooling, Case G1, the near-wall secondary-shock family $S_{r}$ becomes more active, whereas the onset of cellularity occurs farther downstream and $L_{c}$ is smaller. For liquid-kerosene mist cooling, Case L1, the precursor shock $S_{2}$ remains evident, but its wall reflection $S_{2}^{r}$ is noticeably less distinct than in A1. Moreover, only a limited number of reflected waves penetrate across the slip line into the cellular detonation region, indicating that mist cooling weakens the ability of wall-reflected waves to traverse the shear layer. To further quantify the cellular development over the entire investigated injection range, and to facilitate quantitative comparison among different cooling strategies and injection levels, the cellular extent is normalized by the uncooled value $L_{c,A_{0}}$ as $\Lambda_{c}=L_{c}/L_{c,A_{0}}$; the resulting values are summarized in Table~\ref{tab:lambda}.

\begin{figure}[htbp]
\centering
\includegraphics[width=0.65\linewidth]{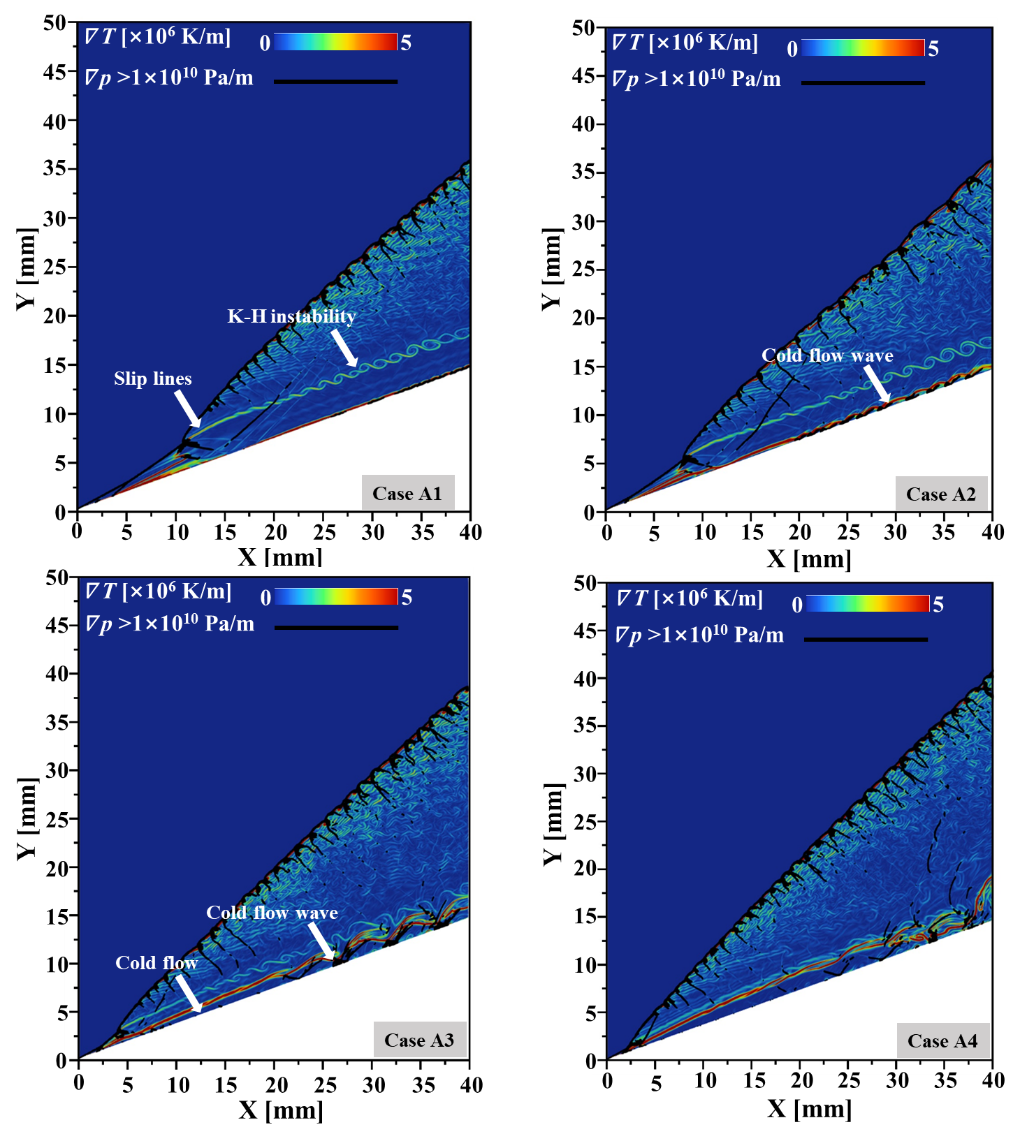}
\caption{Pressure and temperature gradient magnitudes of the cases with air cooling at different $\dot{m}_{c}$.}
\label{fig:tgrad-air}
\end{figure}

For air cooling, $\Lambda_{c}$ increases monotonically from 1.44 to 1.90 with increasing coolant injection, indicating progressively earlier development and an enlarged cellular region. Gaseous-kerosene cooling exhibits a similar increasing trend, but with a weaker response: $\Lambda_{c}$ increases from 0.83 to 1.42, exceeding the uncooled value only at the higher injection levels. In contrast, no monotonic increase is observed for liquid-kerosene mist cooling, for which $\Lambda_{c}$ remains between 0.78 and 1.07 over the investigated range.

Overall, the results suggest that the coolant medium and injection level modulate the downstream cellular development differently. The observed differences are associated with changes in the precursor-shock and reflected-wave structures, as well as their interaction with the slip line and cellular region. Fig.~\ref{fig:tgrad-air} shows the temperature gradient fields for air cooling at different $\dot{m}_{c}$. A pressure gradient threshold contour is overlaid in black, defined by $\left|\nabla p(x,y)\right|>1\times10^{10}\ \mathrm{Pa/m}$, to highlight shocks and associated wave features. The results indicate that the primary ODW structure remains stably wall-attached over the investigated $\dot{m}_{c}$ range, with no macroscopic decoupling between the shock and reaction zones.

As $\dot{m}_{c}$ increases, the triple point shifts upstream, indicating a further upstream displacement of the initiation region. Meanwhile, the near-wall cold layer generated by injection strengthens near the exit, and its interface with the main flow appears as a more distinct high-gradient band in the temperature gradient field; correspondingly, stronger compression waves and localized shocks are induced in the vicinity of the cooling slots. For $\dot{m}_{c}\geq0.009$\,kg/s, the cold layer penetrates into the post-wave region and interacts with the slip line, substantially reducing shear layer fluctuations downstream. K-H structures remain discernible for $\dot{m}_{c}=0.001$--$0.009$\,kg/s, whereas they become essentially indistinguishable at $\dot{m}_{c}=0.015$\,kg/s. Overall, increasing the air injection level both drives the initiation region upstream and suppresses post-wave instabilities through cold layer and shear layer coupling, thereby reshaping the near-wall wave system.

\begin{figure}[htbp]
\centering
\includegraphics[width=0.65\linewidth]{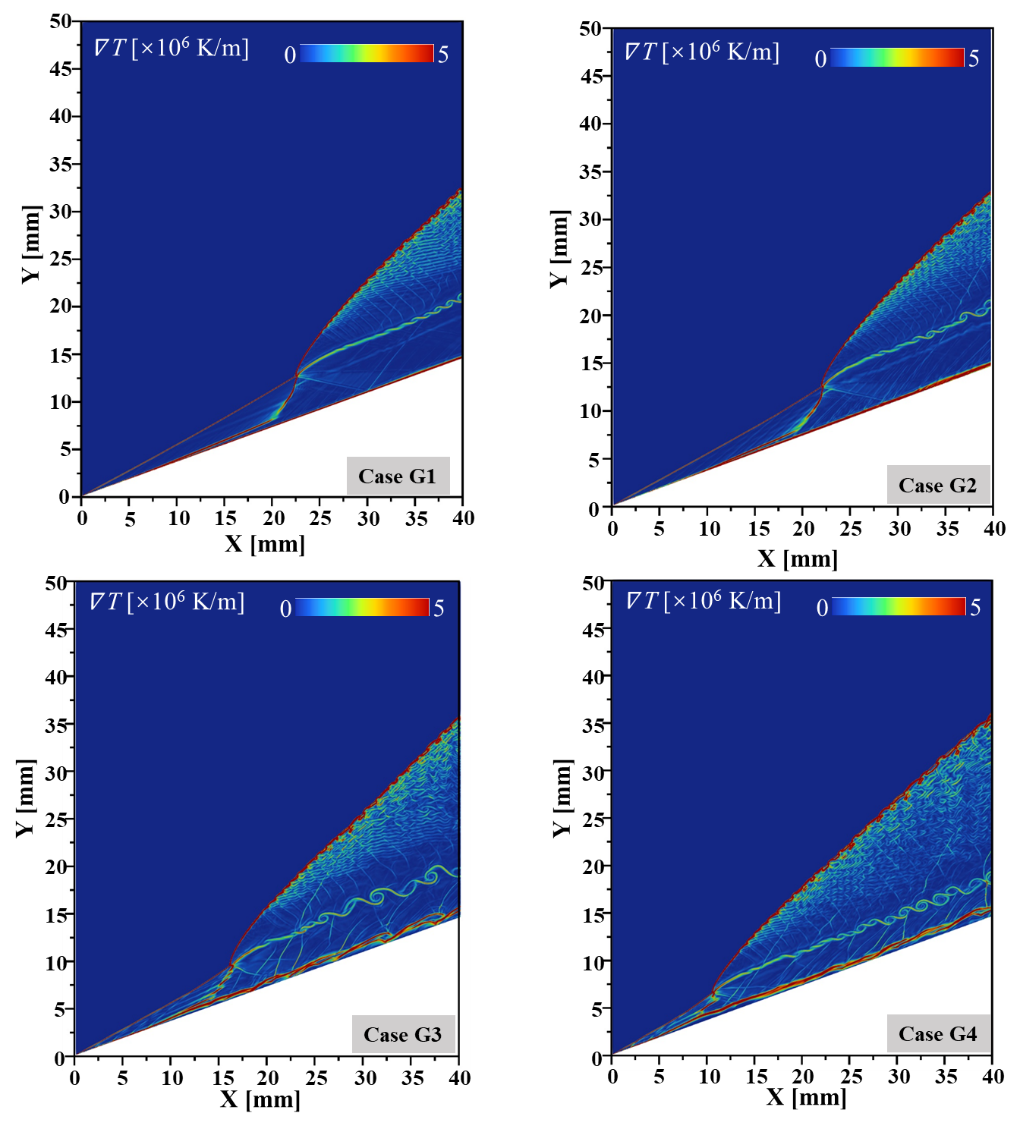}
\caption{Temperature gradient distributions of the gaseous-kerosene cooling cases at different $\dot{m}_{c}$.}
\label{fig:tgrad-gaseous}
\end{figure}

Fig.~\ref{fig:tgrad-gaseous} compares the oblique detonation flow field for gaseous-kerosene cooling at different $\dot{m}_{c}$. In all cases, the primary oblique detonation front remains continuous and wall-attached, and no macroscopic shock-reaction decoupling is observed, indicating that the oblique detonation structure is sustained over the investigated $\dot{m}_{c}$ range. As $\dot{m}_{c}$ increases, the triple point in the initiation region exhibits a modest upstream shift. A more pronounced response is found in the post-wave slip line. At low $\dot{m}_{c}$, the slip line is relatively straight, whereas for $\dot{m}_{c}\geq0.009$\,kg/s it develops clearer quasi-periodic undulations and roll-up structures, indicating an enhancement of shear layer instability.

When $\dot{m}_{c}$ is further increased to $0.015$\,kg/s, the instability persists but its characteristic scale becomes finer. Overall, increasing the gaseous-kerosene injection level preserves the stability of the primary detonation structure while intensifying post-wave shear layer disturbances and altering their evolution, which can in turn affect downstream wave organization and mixing.

\begin{figure}[!htbp]
\centering
\includegraphics[width=0.65\linewidth]{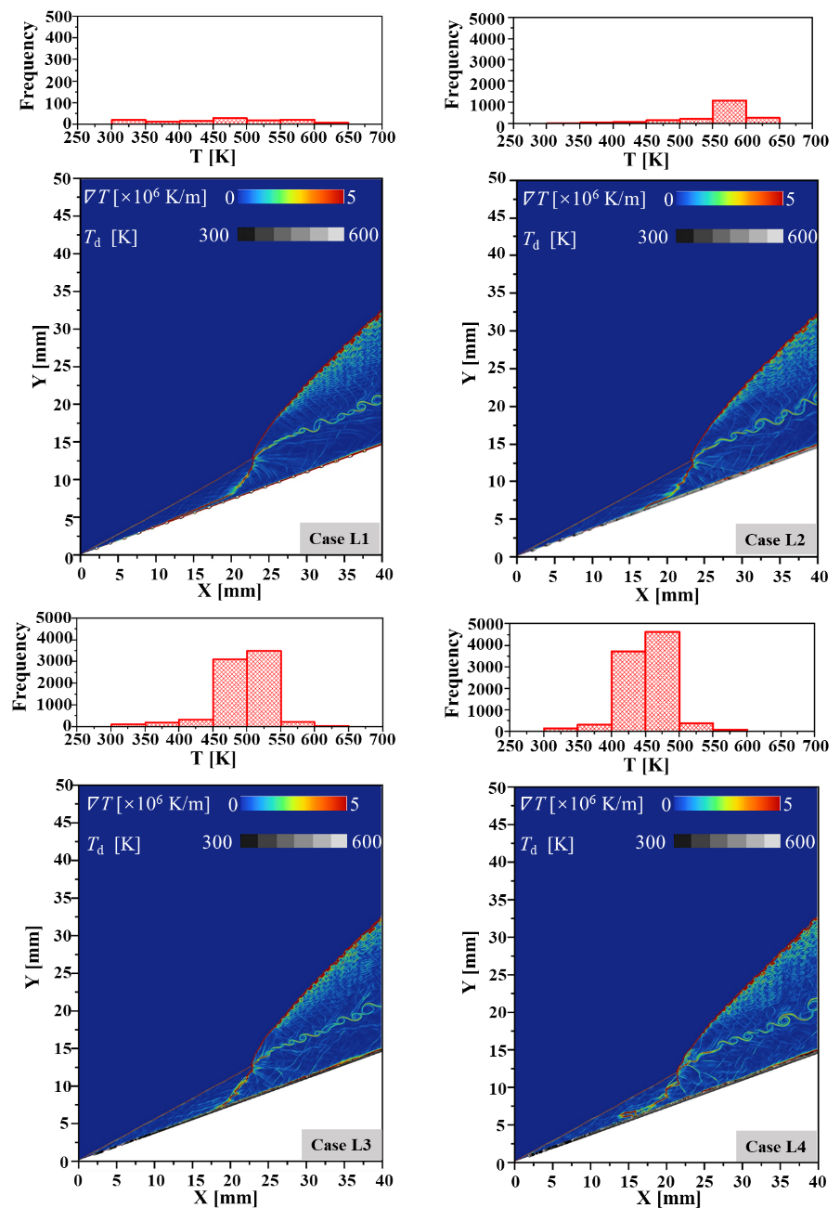}
\caption{Temperature gradient distributions of the liquid-kerosene mist cooling cases at different $\dot{m}_{c}$.}
\label{fig:tgrad-mist}
\end{figure}

Fig.~\ref{fig:tgrad-mist} compares the oblique detonation flow field for liquid-kerosene mist cooling at different $\dot{m}_{c}$. In all cases, the ODW front remains continuous and wall-attached, and the spray primarily modifies the near-wall injection-affected region and the post-wave two-phase mixing and evaporation processes.

The droplet thermal state exhibits a pronounced dependence on $\dot{m}_{c}$. At low $\dot{m}_{c}$, fewer droplets survive within the domain, indicating more rapid heating and evaporation after droplets enter the hot post-wave region. For $\dot{m}_{c}=0.003$--$0.009$\,kg/s, the droplet temperature distribution becomes more concentrated, suggesting a more uniform heating history. When $\dot{m}_{c}$ is further increased to $0.015$\,kg/s, the distribution extends toward lower temperatures and becomes more dispersed, implying stronger near-wall cooling and a thicker spray layer that promote a clearer separation of droplet thermal states.

Together with the wave-front location metric $x_{f}(y)$ introduced in Fig.~\ref{fig:pressure-coolants}, the results highlight distinct impacts of coolant medium on detonation anchoring. Air cooling most readily drives an upstream shift of the wave front, whereas gaseous-kerosene cooling exhibits a more pronounced upstream shift only at higher $\dot{m}_{c}$. In contrast, mist cooling produces no appreciable overall upstream migration over the investigated range. Overall, mist cooling provides near-wall cooling and post-wave modulation while perturbing the geometric anchoring of the initiation region only weakly, which is potentially advantageous for maintaining global flow-field stability. A comprehensive assessment still requires consideration of near-wall temperature metrics and related thermal protection indicators.

\subsection{Near-wall thermal response and cooling effectiveness}

\begin{figure}[!htbp]
\centering
\includegraphics[width=0.35\linewidth]{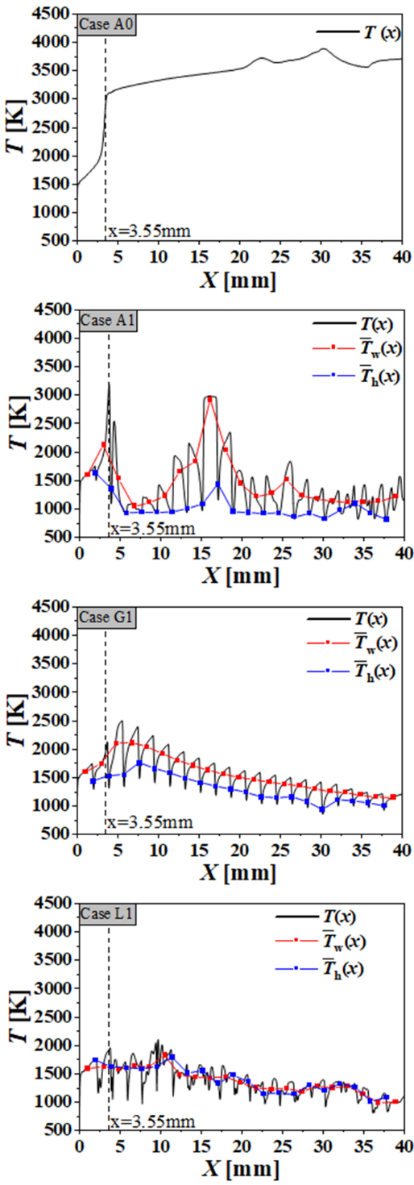}
\caption{Streamwise variation of the near-wall fluid temperature for the no-cooling case and the cases with $\dot{m}_{c}=0.001$\,kg/s.}
\label{fig:tw-baseline}
\end{figure}

Fig.~\ref{fig:tw-baseline} presents the near-wall fluid temperature distribution $T(x)$ sampled along the wall for the uncooled case and the three cooling strategies at $\dot{m}_{c}=0.001$\,kg/s. For the three injection-cooled strategies, the temperature is further phase-averaged over each injection period to obtain the mean temperature along the interval between adjacent injections, $\bar{T}_{w}(x)$, and the mean temperature along the injection segment, $\bar{T}_{h}(x)$. This decomposition enables the spatially periodic thermal response induced by the discrete coolant injection arrangement to be distinguished from the overall film-coverage and downstream recovery behavior. Because the geometric discontinuity near the wedge tip leads to locally elevated $y^{+}$ and may bias the near-wall temperature, the following comparisons are restricted to the region $x\geq x_{0}$, where $x_{0}=3.55$\,mm is the effective sampling start.

Compared with the uncooled baseline, all three cooling strategies significantly reduce the near-wall temperature even at $\dot{m}_{c}=0.001$\,kg/s, confirming that injection cooling can effectively mitigate the wall-adjacent thermal environment. The cooling signatures, however, differ substantially among the coolant media. Case A1 remains strongly non-uniform in the streamwise direction and exhibits pronounced downstream reheating with sharp temperature spikes, with peak values reaching about 3000 K. This behavior is consistent with the upstream shift of the detonation structure and the enhanced near-wall compression and reflected-wave activity, which promote localized reheating. Case G1 shows a clear streamwise injection-periodic near-wall temperature response. The mean temperature over the slot segment is systematically lower than that over the streamwise wall segment between adjacent cooling slots, and $T(x)$ displays a characteristic slow-rise and sharp-drop sawtooth pattern, indicating progressive heating and decay of the cooling film downstream of each slot followed by re-establishment of a fresh low-temperature layer at the next slot. Case L1 provides the strongest temperature suppression and the best streamwise uniformity among the three strategies. The difference between the slot-averaged and inter-slot-averaged temperatures is the smallest, and the peak temperature remains below 2100 K throughout, well below the peaks observed in A1 and G1. These results indicate that evaporative cooling, phase-change heat absorption, and two-phase mixing more effectively weaken the hot post-wave region and suppress slot-scale hot-spot formation.

\begin{figure}[!htbp]
\centering
\includegraphics[width=0.95\linewidth]{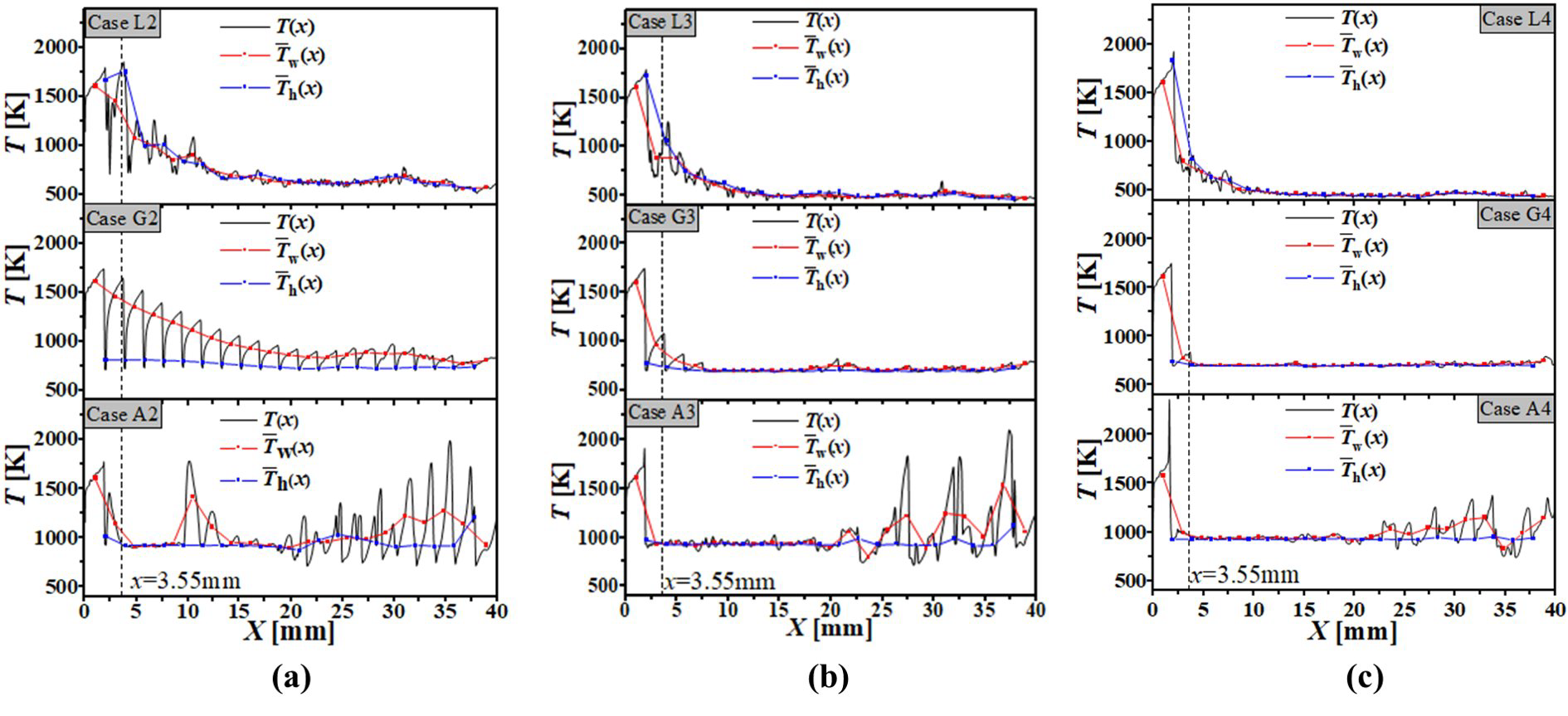}
\caption{Streamwise variations of the near-wall fluid temperature under different $\dot{m}_{c}$ and coolant media. (a) $\dot{m}_{c}=0.003$\,kg/s; (b) $\dot{m}_{c}=0.009$\,kg/s; (c) $\dot{m}_{c}=0.015$\,kg/s.}
\label{fig:tw-mass-ratios}
\end{figure}

Fig.~\ref{fig:tw-mass-ratios}(a--c) compares the near-wall temperature distributions along the wall for the three coolant media at $\dot{m}_{c}$ = 0.003, 0.009, and 0.015\,kg/s. The black curve is the pointwise sampled temperature $T(x)$. The red and blue dotted curves denote the period-averaged temperatures over the streamwise wall segment between adjacent cooling slots and the slot segment, respectively. The vertical dashed line at $x=3.55$\,mm marks the effective sampling start where $y^{+}<5$. The degree of convergence between the slot-averaged and inter-slot-averaged temperatures is used to indicate the continuity of post-slot wall attachment; a smaller separation implies weaker decay of cooling between slots and more complete coverage.

Overall, liquid-kerosene mist cooling provides the strongest near-wall fluid temperature suppression and the best streamwise uniformity among the cooled cases. At $\dot{m}_{c}=0.003$\,kg/s, $\bar{T}_{w}(x)$ and $\bar{T}_{h}(x)$ rapidly converge over most of the downstream region, and the temperature can be maintained at about 600 K, indicating that evaporative heat absorption and two-phase mixing markedly weaken local streamwise near-wall temperature peaks and inter-slot temperature recovery.

Gaseous-kerosene film cooling is the next most effective. At $\dot{m}_{c}=0.003$\,kg/s it still exhibits a pronounced streamwise injection-periodic peak--valley response, whereas at $\dot{m}_{c}=0.009$ and $0.015$\,kg/s the two averaged temperatures tend to converge, indicating more continuous streamwise persistence of the near-wall cooled region. The lowest downstream temperatures, however, are limited by the coolant inlet temperature and the sensible-heat capacity of the gas phase, giving an overall level of about 700 K.

Air cooling shows the strongest non-uniformity at the same $\dot{m}_{c}$. Although increasing $\dot{m}_{c}$ reduces the mean temperature, sharp streamwise spikes and enhanced downstream fluctuations persist, and the convergence between slot and inter-slot averages remains weaker than in Cases G and L. This behavior suggests that the air-cooling layer is more susceptible to disturbances from the wave system and cold-flow structures, leading to intermittent reheating and reduced streamwise persistence of the near-wall cooling layer.

To enable a consistent comparison of cooling performance across coolant media, a dimensionless adiabatic cooling effectiveness is defined for $x\geq x_{0}$ by referencing the uncooled adiabatic wall temperature $T_{\mathrm{aw},0}(x)$ and normalizing with the coolant inlet temperature

\begin{equation}
\eta_{T}(x)=\frac{T_{\mathrm{aw},0}(x)-T_{\mathrm{aw}}(x)}{T_{\mathrm{aw},0}(x)-T_{c,\mathrm{in}}}
\label{eq:eta}
\end{equation}
Here, $T_{\mathrm{aw}}(x)$ denotes the gas-side adiabatic wall temperature extracted from the wedge-wall boundary, and $T_{\mathrm{aw},0}(x)$ is the corresponding value for the uncooled case A0. Since the wedge wall is treated as adiabatic, the present thermal assessment is based on $T_{\mathrm{aw}}(x)$, while solid-wall heat conduction and non-adiabatic wall heat transfer are beyond the scope of the present model. $T_{c,\mathrm{in}}$ denotes the coolant inlet temperature. Specifically, $T_{c,\mathrm{in}}=929.8\ \mathrm{K}$ for air cooling, $T_{c,\mathrm{in}}=700\ \mathrm{K}$ for gaseous-kerosene cooling, and $T_{c,\mathrm{in}}=300\ \mathrm{K}$ for liquid-kerosene mist cooling.

\begin{figure}[!htbp]
\centering
\includegraphics[width=0.35\linewidth]{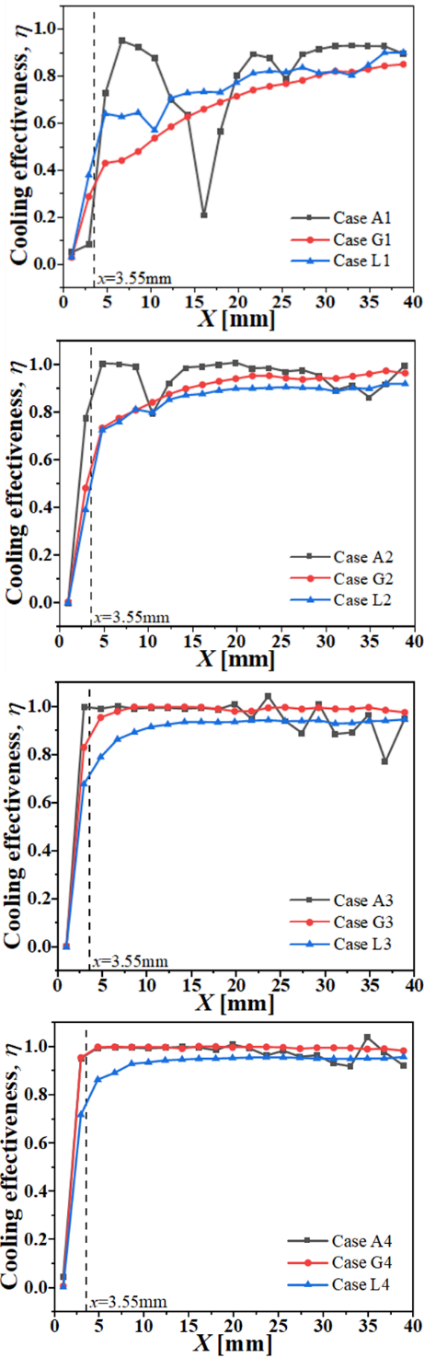}
\caption{Mean cooling effectiveness over the streamwise wall segment between adjacent cooling slots for different cases.}
\label{fig:eta}
\end{figure}

Fig.~\ref{fig:eta} summarizes the mean cooling effectiveness $\eta_{T}(x)$ over the wall segment between adjacent cooling slots for different coolant media and mass flow rates.

At $\dot{m}_{c}=0.001$\,kg/s, Case A1 attains relatively high local effectiveness in some regions, but exhibits pronounced fluctuations over a streamwise interval. Consistent with the wave-structure analysis based on the pressure-gradient field, this interval corresponds to frequent wall impingement of the precursor shock family $S_{2}$ and intensified activity of its reflected waves. The associated shock--wall interactions produce localized compression heating, which manifests as abrupt drops and strong oscillations in $\eta_{T}$. Thus, although A1 can be effective locally, it is more sensitive to wave unsteadiness and yields poorer streamwise robustness. In contrast, Cases G1 and L1 display much smoother $\eta_{T}(x)$ distributions at the same $\dot{m}_{c}$, indicating weaker amplification of wave disturbances in the near-wall thermal environment. The effectiveness for L1 is mostly concentrated in the range 0.6--0.9, whereas G1 spans approximately 0.4--0.85.

When the coolant mass flow rate is increased to $\dot{m}_{c}=0.003$\,kg/s, the effectiveness increases for all three media. The oscillations in A2 are reduced, but downstream variations associated with the cold-flow wave structure remain visible. By comparison, G2 and L2 reach a stable high-effectiveness plateau for $x\geq 5$\,mm, with $\eta_{T} \approx 0.73$--$0.97$ for G2 and $\eta_{T} \approx 0.73$--$0.92$ for L2, and the gap between the two fuel-based strategies further narrows. This behavior indicates that increasing the fuel-based injection level promotes a more continuous wall-attached protective layer.

For higher injection levels, $\dot{m}_{c}=0.009$ and $0.015$\,kg/s, the upstream thermal protection is further strengthened. The upstream oscillations for air cooling largely vanish, but the downstream effectiveness remains modulated by the cold-flow wave structure, being more pronounced at $\dot{m}_{c}=0.009$\,kg/s and partially alleviated at $\dot{m}_{c}=0.015$\,kg/s. In addition, mist cooling no longer shows a clear advantage over gaseous-kerosene cooling in terms of $\eta_{T}$ at high $\dot{m}_{c}$. This trend follows from the normalization in the definition of $\eta_{T}$. The lower spray inlet temperature increases the denominator, such that $\eta_{T}$ reflects the utilization of the available coolant-to-baseline temperature difference rather than the absolute temperature reduction alone. Consequently, the choice of coolant inlet temperature $T_{c}$ affects the quantitative comparison of $\eta_{T}$ among different coolant media. Nevertheless, the principal cooling trends remain unchanged when the absolute near-wall fluid temperature distributions are examined independently. In terms of absolute near-wall fluid temperature, mist cooling still provides stronger temperature suppression than gaseous-kerosene cooling.

\begin{figure}[!htbp]
\centering
\includegraphics[width=0.8\linewidth]{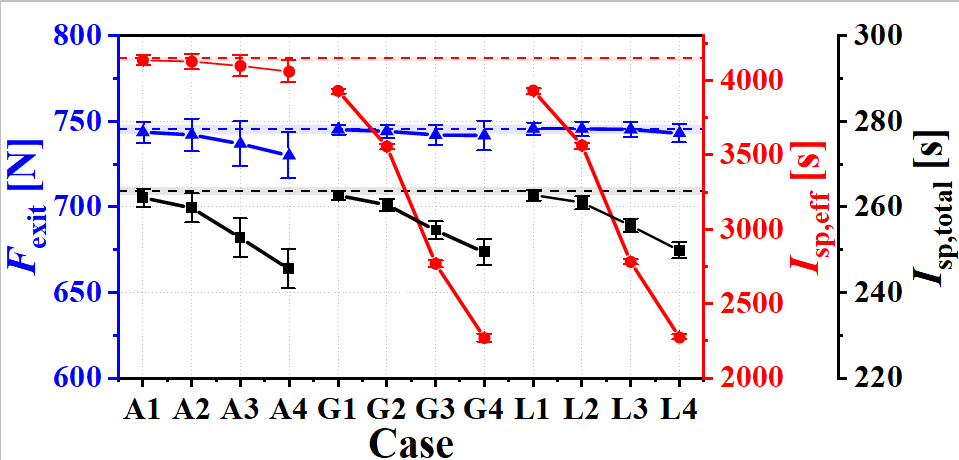}
\caption{Specific impulse and thrust for different cases (error bars and shaded bands indicate $\pm1$ standard deviation; dashed lines indicate the no-cooling baseline).}
\label{fig:thrust}
\end{figure}

Fig.~\ref{fig:thrust} summarizes the trends of exit thrust $F_{\mathrm{exit}}$, fuel-based specific impulse $I_{\mathrm{sp,eff}}$, and total specific impulse $I_{\mathrm{sp,total}}$ for the different cases. The streamwise exit force is evaluated as

\begin{equation}
F_{\mathrm{exit}}=\int_{A_{e}}\left[\rho U_{x}^{2}+(p-p_{b})\right]\mathrm{d}A
\label{eq:thrust}
\end{equation}
where $A_{e}$ is the exit-plane area, $p_{b}=101325\ \mathrm{Pa}$ is the back pressure, and $g$ is the gravitational acceleration. For the two-dimensional simulations, a spanwise thickness of $b=1$\,mm is adopted when evaluating the dimensional exit force and mass flow rates. The fuel-based and total specific impulses are defined as

\begin{equation}
I_{\mathrm{sp,eff}}=\frac{F_{\mathrm{exit}}}{g\dot{m}_{f,\mathrm{tot}}}
\label{eq:ispeff}
\end{equation}

\begin{equation}
I_{\mathrm{sp,total}}=\frac{F_{\mathrm{exit}}}{g(\dot{m}_{\mathrm{main}}+\dot{m}_{c})}
\label{eq:isptotal}
\end{equation}
Here,
\begin{equation}
\dot{m}_{f,\mathrm{tot}}=\begin{cases}\dot{m}_{f,\mathrm{main}}, & \text{for A0 and A1--A4}\\[2pt]\dot{m}_{f,\mathrm{main}}+\dot{m}_{c}, & \text{for G and L series}\end{cases}
\label{eq:mftot}
\end{equation}
because the cooling stream in the G- and L-series consists of additional kerosene. The same control-volume boundaries are used for all cases.

The reported $F_{\mathrm{exit}}$ represents the axial force evaluated at the combustor exit. It does not include the inlet momentum contribution, external drag, or the power and losses associated with coolant/fuel delivery; therefore, the corresponding specific impulses are used as exit-force-based propulsion-performance indicators rather than net system-level metrics.

The propulsion-performance parameters are time-averaged over $t\geq0.1$ ms, with the error bars representing $\pm1$ standard deviation of the temporal fluctuations. The dashed lines denote the time-averaged uncooled baseline (Case A0), with the shaded bands indicating the corresponding $\pm1$ standard deviation. The relative standard deviations remain below 1.90\% for all cases, indicating limited temporal fluctuations over the selected averaging interval. Overall, increasing the $\dot{m}_{c}$ leads to reductions in both thrust and specific impulse for all three cooling strategies, indicating an unavoidable momentum and energy penalty associated with injection cooling. The loss mechanisms include added injected mass, mixing losses, and changes in the wave system that reduce the effective expansion capability.

Differences among coolant media are relatively small in terms of $I_{\mathrm{sp,total}}$, suggesting that once the injected mass is consistently included in the denominator, the overall efficiency gap among strategies is substantially compressed. Thrust, by contrast, is more sensitive to the coolant medium. At the same coolant mass flow rate, air cooling produces a more pronounced thrust penalty, consistent with its stronger perturbation of the primary wave system. Fuel-based cooling, either gaseous kerosene or liquid mist, has a smaller impact on thrust at low to moderate coolant mass flow rates. In particular, gaseous-kerosene cooling at $\dot{m}_{c}=0.001$\,kg/s and mist cooling over $\dot{m}_{c}=0.001$--$0.009$\,kg/s yield thrust values closer to the uncooled baseline.

Air cooling does not increase the fuel flow rate and therefore yields a higher numerical value of $I_{\mathrm{sp,eff}}$. For a more comparable system-level assessment, the present discussion emphasizes joint trends in $F_{\mathrm{exit}}$ and $I_{\mathrm{sp,total}}$, while retaining $I_{\mathrm{sp,eff}}$ as a reference measure of fuel utilization. Considering wall thermal protection together with propulsion performance, fuel-based cooling provides a more stable wall-attached protection at low to moderate coolant mass flow rates with a smaller performance penalty. Mist cooling offers an advantageous compromise over $\dot{m}_{c}=0.003$--$0.009$\,kg/s, where strong absolute near-wall fluid temperature suppression is achieved with acceptable performance loss. Air cooling may reach high local instantaneous effectiveness, but it is more sensitive to wave unsteadiness, which promotes downstream fluctuations in cooling performance and a larger thrust penalty. Mist cooling remains most favorable in terms of absolute near-wall fluid temperature control, whereas its normalized effectiveness should be interpreted in conjunction with the absolute temperature level.

\section*{Conclusions}

This study numerically investigates thermal-protection strategies for a kerosene--air ODE combustor operating at a flight Mach number of 10 and an altitude of 15 km. A validated two-dimensional numerical framework is employed to compare air film cooling, gaseous-kerosene film cooling, and liquid-kerosene mist cooling over $\dot{m}_{c}=0.001$--$0.015$\,kg/s. Their effects on the detonation-wave structure, near-wall thermal environment, and propulsion performance are systematically evaluated. The main conclusions are summarized as follows:

(1) The ODW system remains self-sustained for all investigated cooling strategies and $\dot{m}_{c}$, demonstrating robustness to coolant injection within the present operating range. Nevertheless, the local wave response depends strongly on the coolant medium and injection mode. Air injection produces the strongest disturbance to the initiation and triple-point regions and leads to more pronounced modification of the wave system as the $\dot{m}_{c}$ increases. Fuel-based cooling interacts more moderately with the reacting core flow, with gaseous-kerosene and liquid-kerosene mist injection better preserving the baseline detonation structure, particularly at low to moderate $\dot{m}_{c}$.

(2) The three strategies exhibit distinct near-wall cooling characteristics. Air cooling forms a low-temperature near-wall layer but is accompanied by stronger wave-induced thermal fluctuations. Gaseous-kerosene cooling develops a wall-attached protective film whose continuity improves with increasing coolant injection. Liquid-kerosene mist cooling combines droplet evaporation and phase-change heat absorption with near-wall transport and mixing, resulting in strong suppression of the near-wall fluid temperature and a relatively uniform thermal field. At low to moderate $\dot{m}_{c}$, mist cooling achieves the lowest absolute near-wall fluid temperatures among the investigated strategies. At higher injection levels, the continuity of the gaseous-kerosene film improves substantially. Because cooling effectiveness is normalized by the coolant inlet temperature, comparisons based on normalized effectiveness alone may obscure the absolute thermal advantage of the lower-temperature mist; therefore, normalized effectiveness and absolute near-wall fluid temperature should be considered together.

(3) Increasing the $\dot{m}_{c}$ introduces propulsion penalties for all three strategies. Air cooling causes the largest thrust degradation, consistent with its stronger disturbance of the primary wave system, whereas fuel-based cooling better preserves propulsion performance. Considering near-wall fluid temperature suppression, flow-field stability, and propulsion-performance loss together, liquid-kerosene mist cooling provides a favorable balance between thermal protection and propulsion performance over $\dot{m}_{c}=0.003$--$0.009$\,kg/s under the investigated conditions. At $\dot{m}_{c}=0.009$\,kg/s, the average near-wall fluid temperatures between adjacent cooling slots and over the cooling-slot regions are reduced by 86.58\% and 86.66\%, respectively, relative to the uncooled case, while the total specific impulse decreases by 3.02\%; at $\dot{m}_{c}=0.015$\,kg/s, the corresponding reduction in the total specific impulse is 5.32\%. Further increasing $\dot{m}_{c}$ yields only a marginal additional thermal benefit, whereas the propulsion penalty continues to grow. Gaseous-kerosene film cooling also shows potential at higher $\dot{m}_{c}$, where improved streamwise persistence of the near-wall cooling layer can provide sustained near-wall protection with a relatively moderate propulsion penalty.

Finally, we would like to note that the above findings should be interpreted within the scope of the present study. The conclusions are drawn for a single $20^\circ$ wedge geometry at a flight Mach number of 10 and an altitude of 15 km, and the simulations solve the Navier--Stokes equations without an explicit turbulence model. Moreover, the cooling performance is assessed through the near-wall fluid temperature and the cooling effectiveness under adiabatic-wall conditions, without conjugate heat transfer or direct prediction of the actual wall temperature and heat flux. From a practical perspective, issues such as the fuel-vaporization or high-pressure-air supply requirements, the possible fouling or blockage of the small injection passages, and system-level integration, together with experimental validation under broader operating conditions and geometric configurations, are left for future work.

\bibliographystyle{elsarticle-num}
\bibliography{refs}

\end{document}